\documentclass[epj]{svjour}
\usepackage{graphics}
\usepackage{amsmath,amssymb}
\usepackage{graphicx}
\usepackage[dvips]{epsfig}
\usepackage{subfigure}

\newcommand{\pol}[1]{\mathaccent"017E{#1}}

\newcommand{\dd}{\mbox{\rm d}}
\newcommand{\dpce}{\mbox{$dp\to \{pp\}_{\!s}n$}}
\newcommand{\dpcepol}{\mbox{$\pol{d}p\to \{pp\}_{\!s}n$}}
\newcommand{\dpncep}{\mbox{$\pol{d}\pol{p}\to \{pp\}_{\!s}n$}}
\def\fmn#1#2{\mbox{${\textstyle \frac{#1}{#2}}$}}

\newcommand{\half}{\mbox{${\textstyle \frac{1}{2}}$}}

\newcommand{\Szero}{\mbox{$^{1\!}S_0$}}

\newcommand{\dpdppi}{$dp \rightarrow p_{\rm sp}d\pi^{0}$}

\newcommand{\npdpi}{$np \rightarrow d\pi^{0}$}
\newcommand{\ppdpi}{$pp \rightarrow d\pi^{+}$}

\begin{document}
\title{The neutron-proton charge-exchange amplitudes measured in the $\boldsymbol{dp \to ppn}$ reaction}
\author{D.~Mchedlishvili\inst{1,2}\and
 S.~Barsov\inst{3}\and
 J.~Carbonell\inst{4}\and
 D.~Chiladze\inst{1,2}\and
 S.~Dymov\inst{5,6}\and
 A.~Dzyuba\inst{3} \and
 R.~Engels\inst{2}\and
 R.~Gebel\inst{2}\and
 V.~Glagolev\inst{7}\and
 K.~Grigoryev\inst{2,3}\and
 P.~Goslawski\inst{8}\and
 M.~Hartmann\inst{2}\and
 A.~Kacharava\inst{2}\and
 V.~Kamerdzhiev\inst{2}\and
 I.~Keshelashvili\inst{1,9}\and
 A.~Khoukaz\inst{8}\and
 V.~Komarov\inst{5}\and
 P.~Kulessa\inst{10}\and
 A.~Kulikov\inst{5}\and
 A.~Lehrach\inst{2}\and
 N.~Lomidze\inst{1}\and
 B.~Lorentz\inst{2}\and
 G.~Macharashvili\inst{1,5}\and
 R.~Maier\inst{2}\and
 S.~Merzliakov\inst{2,5}\and
 M.~Mielke\inst{8}\and
 M.~Mikirtychyants\inst{2,3}\and
 S.~Mikirtychyants\inst{2,3}\and
 M.~Nioradze\inst{1}\and
 H.~Ohm\inst{2}\and
 M.~Papenbrock\inst{8}\and
 D.~Prasuhn\inst{2}\and
 F.~Rathmann\inst{2}\and
 V.~Serdyuk\inst{2}\and
 H.~Seyfarth\inst{2}\and
 H.J.~Stein\inst{2}\and
 E.~Steffens\inst{6}\and
 H.~Stockhorst\inst{2}\and
 H.~Str\"oher\inst{2}\and
 M.~Tabidze\inst{1}\and
 S.~Trusov\inst{11}\and
 Yu.~Uzikov\inst{5,12}\and
 Yu.~Valdau\inst{2,3}\and
 C.~Wilkin\inst{13}\thanks{e-mail: c.wilkin@ucl.ac.uk} }
\authorrunning{D.~Mchedlishvili \emph{et.al.}}
\titlerunning{The neutron-proton charge-exchange amplitudes measured in the $dp \to ppn$
reaction}
%
\institute{High Energy Physics Institute, Tbilisi State University, GE-0186
Tbilisi, Georgia
\and  Institut f\"ur Kernphysik and J\"ulich Centre for Hadron Physics, Forschungszentrum J\"ulich, D-52425
J\"ulich, Germany
\and High Energy Physics Department, Petersburg Nuclear Physics
Institute, RU-188350 Gatchina, Russia
\and  Institut de Physique Nucl\'{e}aire, Universit\'e Paris-Sud, IN2P3-CNRS, F-91406 Orsay Cedex, France
\and Laboratory of Nuclear Problems, JINR, RU-141980 Dubna, Russia
\and  Physikalisches Institut II, Universit\"at Erlangen--N\"urnberg,
D-91058 Erlangen, Germany
\and Laboratory of High Energies, JINR, RU-141980 Dubna, Russia
\and Institut f\"ur Kernphysik, Universit\"at M\"unster, D-48149
M\"unster, Germany
\and Department of Physics, University of Basel, Klingelbergstrasse
82, CH-4056 Basel, Switzerland
\and H.~Niewodnicza\'{n}ski Institute of Nuclear Physics PAN, PL-31342
Krak\'{o}w, Poland
\and Institut f\"ur Kern- und Hadronenphysik,
Forschungszentrum Rossendorf, D-01314 Dresden, Germany
\and Department of Physics, M.~V.~Lomonosov Moscow State
University, RU-119991 Moscow, Russia
\and Physics and Astronomy Department, UCL, Gower Street, London,
WC1E 6BT, UK}
\date{Received: \today / Revised version:}
\abstract{The unpolarised differential cross section and the two deuteron
tensor analysing powers $A_{xx}$ and $A_{yy}$ of the \dpcepol\
charge-exchange reaction have been measured with the ANKE spectrometer at the
COSY storage ring. Using deuteron beams with energies 1.2, 1.6, 1.8, and
2.27~GeV, data were obtained for small momentum transfers to a $\{pp\}_{\!s}$
system with low excitation energy. The results at the three lower energies
are consistent with impulse approximation predictions based upon the current
knowledge of the neutron-proton amplitudes. However, at 2.27~GeV, where these
amplitudes are far more uncertain, agreement requires a reduction in the
overall double-spin-flip contribution, with an especially significant effect
in the longitudinal direction. These conclusions are supported by
measurements of the deuteron-proton spin-correlation parameters $C_{x,x}$ and
$C_{y,y}$ that were carried out in the \dpncep\ reaction at 1.2 and 2.27~GeV.
The values obtained for the proton analysing power $A_y^p$ also suggest the
need for a radical re-evaluation of the neutron-proton elastic scattering
amplitudes at the higher energy. It is therefore clear that such measurements
can provide a valuable addition to the neutron-proton database in the
charge-exchange region.
} 
\maketitle

%
%

\section{Introduction}
\label{sec1}

An understanding of the nucleon-nucleon ($NN$) interaction is fundamental for
the whole of nuclear and hadronic physics. The SAID database and analysis
program~\cite{ARN2000} have proved to be truly invaluable tools over many
years for researchers working in this area. The general procedure adopted
here is to take all the $NN$ elastic scattering data in the literature in
order to perform a phase shift analysis up to a certain orbital angular
momentum $L_{\textrm{max}}$ and use a theoretical model for higher $L$. When
significant new experimental data become available, the value of
$L_{\textrm{max}}$ can be increased. By assuming that the phase shifts vary
smoothly with beam energy, predictions can be made for observables at a
particular energy and it is in this way that the SAID program is most
commonly used.

Clearly any amplitude analysis can only be as good as the data used in its
implementation. Though lots of proton-proton observables have been measured
up to high energies, there are significant gaps in our knowledge for beam
energies $T_N \gtrsim 2$~GeV, especially at small angles. The situation is
even more serious for neutron-proton scattering where there are major holes
in the experimental database above about 1~GeV and data that do exist are not
necessarily very well reproduced by the SAID program. For example, the only
differential cross section data for large angle $np$
scattering~\cite{BIZ1975}, in the so-called charge-exchange region, seem to
be consistently over-predicted in the SAID analysis.

Extra information in the $np$ sector can be found by performing experiments
with a deuteron beam or a deuterium target. The simplest of these involves
measuring the ratio of the forward charge-exchange cross section of a neutron
on a deuterium target to that on hydrogen,
\begin{equation}
R_{np}(0)=\left.\frac{\dd\sigma(nd\to pnn)/\dd{t}} {\dd\sigma(np\to
pn)/\dd{t}}\right|_{\theta=0}, \label{Rnp}
\end{equation}
where $t$ is the square of the four-momentum transfer between the initial
neutron and final proton. Due to the Pauli principle, when the two final
neutrons are in a relative $S$-wave their spins must be antiparallel and the
system is in the \Szero\ state. Under such circumstances the $nd\to p\{nn\}$
reaction involves a spin flip from the $S=1$ of the deuteron to the $S=0$ of
the dineutron. In impulse approximation the ratio determines
\begin{equation}
R_{np}(0)=\frac{2}{3}\left(\frac{\sigma_{\textit{sf}}}
{\sigma_{\textit{sf}}+\sigma_{\textit{nsf}}}\right),
\label{Rnp2}
\end{equation}
where $\sigma_{\textit{sf}}$ ($\sigma_{\textit{nsf}}$) is the forward $np$
charge-exchange differential cross section with (without) a
spin-flip~\cite{DEA1972}.

Extensive measurements of $R_{np}(0)$ have been reported and these have
recently been extended up to 2~GeV at Dubna~\cite{SHA2009}. Although impulse
approximation predictions for $R_{np}(0)$ on the basis of the current SAID
amplitudes reproduce well these observations from 300~MeV up to about
800~MeV, there are serious discrepancies above 1~GeV~\cite{LEH2010}. In terms
of Eq.~\eqref{Rnp2}, it seems that the SAID solution overestimates the
contribution of the spin-flip amplitudes to the forward $np$ charge-exchange
cross section.

More detailed information on the $np$ charge-exchange amplitudes can be
derived by using a polarised deuteron beam or target and studying the
\dpcepol\ reaction~\cite{BUG1987}. To achieve maximum sensitivity, the
excitation energy $E_{pp}$ in the final $pp$ system must be very low so that
the final diproton is in the \Szero\ configuration. Experiments from a few
hundred MeV up to 2~GeV~\cite{ELL1987,KOX1993} have generally borne out well
predictions based upon the impulse approximation~\cite{BUG1987}.

In order to constrain the $np$ amplitudes using this approach, the ANKE
collaboration has embarked on a systematic programme to measure the \dpcepol\
observables up to the maximum COSY deuteron energy of $T_d\approx
2.3$~GeV~\cite{KAC2005}. The proof of principle of the method was the
experiment carried out at a deuteron energy of 1.17~GeV where, because of the
wealth of neutron-proton data, the SAID amplitudes used as input in the
calculations should be quite reliable. The measured values of the unpolarised
cross section and the two deuteron Cartesian tensor analysing powers $A_{xx}$
and $A_{yy}$ were then quantitatively reproduced in impulse
approximation~\cite{CHI2009}. Dilutions of the signals due to higher partial
waves in the final $pp$ system were taken into account in the
calculations~\cite{CAR1991}.

The results of similar measurements are presented here at deuteron beam
energies of 1.2, 1.6, 1.8, and 2.27~GeV. Whereas the unpolarised differential
cross section is correctly predicted in impulse approximation at the three
lower energies, this gives a result that is about 15\% too high at 2.27~GeV.
Such a factor is in fact consistent with the discrepancy between the SAID
$np$ predictions and the data in the charge-exchange region~\cite{BIZ1975},
taking into account the SAID overestimation of the spin-flip contributions,
as measured in the inclusive $R_{np}(0)$ experiment~\cite{SHA2009,LEH2010}.

The situation is rather similar for $A_{xx}$ and $A_{yy}$, where impulse
approximation reproduces the measurements very well at 1.8~GeV and below but
gives a much poorer description at 2.27~GeV. Agreement can be restored if the
spin-longitudinal amplitude is reduced by $\approx 25$\% compared to the SAID
values. This is not entirely unexpected because the deuteron tensor
analysing power in the forward direction is closely linked to the
longitudinal and transverse spin-transfer coefficients in $np$ charge
exchange. Since there are no measurements of these parameters at small angles
in the $T_n \approx 1.135$~GeV region, the SAID amplitudes are fixed here
mainly by the assumptions in the model.

Some confirmation of the renormalisation of the spin-longitudinal amplitude
is offered by the values of the two deuteron-proton spin-correlation
parameters $C_{x,x}$ and $C_{y,y}$ that were measured at 1.2 and 2.27~GeV by
replacing the hydrogen cluster target with a specially designed cell that can
be filled with  polarised gas~\cite{GRI2007}. In the \Szero\ limit these
parameters depend upon interferences between the spin-spin amplitudes and,
although the SAID input reproduces well the 1.2~GeV data, at 2.27~GeV a 25\%
reduction in the spin-longitudinal term leads to a much better description.

The polarised hydrogen target~\cite{MIK2012} also allowed us to
measure the proton analysing power $A_y^p$ in the $d\pol{p}\to
\{pp\}_{\!s}n$ reaction and the conclusions to be drawn here
are broadly similar. Despite the $A_y^p$ signal being quite
small over the ANKE angular range, it is reproduced
quantitatively in the impulse approximation calculations at
1.2~GeV, though these fail badly at 2.27~GeV. This observable
depends upon the interference of a spin-orbit amplitude with a
spin-spin term and it would seem that the current SAID
analysis~\cite{ARN2000} grossly underestimates the imaginary
part of the spin-orbit amplitude at the higher energy.

We have described the phenomenology of the deuteron charge-exchange reaction
at some length in earlier publications~\cite{CHI2009,CHI2006a,CHI2006b} but,
in order to make the current paper more self-contained, some of this material
is repeated in sect.~\ref{sec2}. The experimental set-up for measuring with
the hydrogen cluster-jet target is identical to that used in our earlier
work~\cite{CHI2009} and so most of the emphasis in sect.~\ref{sec3} is on the
polarised hydrogen gas cell used in the spin-correlation measurements. The
measurements of the luminosity through the observation of the quasi-free
$pn\to d\pi^0$ reaction are reported in sect.~\ref{sec4}, where the results
for the unpolarised \dpce\ differential cross sections are compared with
impulse approximation calculations.

The polarisation of the deuteron beam was established at 1.2~GeV and, since
there are no depolarising resonances for the deuteron in the COSY energy
range, the analysing powers for the \dpcepol\ reaction could be measured at
various energies and the results are presented in sect.~\ref{sec5}. The
polarisation measurements with the gas cell have no parallel in our earlier
work. The method used here relies on the data taken with the polarised
deuterium target and the measurement of the analysing power of quasi-free
$n\pol{p}\to d\pi^0$. Taken together with the measurement of the vector
polarisation of the deuteron beam, this allowed us to extract the
spin-correlation and proton analysing power results given in
sect.~\ref{sec6}. Our conclusions regarding the usefulness of charge exchange
on the deuteron in the study of neutron-proton elastic scattering amplitudes
are drawn in sect.~\ref{sec8}.

%
%
\section{Deuteron charge exchange in impulse approximation}
\label{sec2}\setcounter{equation}{0}

The cross section and spin observables for the \dpce\ reaction have been
extensively discussed in the literature~\cite{BUG1987,CAR1991} and only
essential formulae are collected here. In impulse approximation the deuteron
charge exchange amplitude is proportional to a neutron-proton charge exchange
amplitude times a form factor that reflects the overlap between the initial
deuteron wave function and that of the outgoing diproton system.

The elementary $np\to pn$ amplitude may be written in terms of five scalar
amplitudes in the cm system as
\begin{eqnarray}
\nonumber f_{np}&=&\alpha(q) +i\gamma(q)
(\vec{\sigma}_{1}+\vec{\sigma}_{2})\cdot\vec{n} +\beta(q)
(\vec{\sigma}_{1} \cdot {\bf n})(\vec{\sigma}_{2}\cdot\vec{n})\\
&&+\delta(q)(\vec{\sigma}_{1}\cdot\vec{m})(\vec{\sigma}_{2}\cdot\vec{m})
+\varepsilon(q)(\vec{\sigma}_{1}\cdot\vec{l})(\vec{\sigma}_{2}\cdot\vec{l}),
 \label{fpn}
\end{eqnarray}
where $q=\sqrt{-t}$ is the three-momentum transfer and the Pauli matrices
$\vec{\sigma}$ are sandwiched between neutron and proton spinors. Here
$\alpha$ is the spin-independent amplitude between the initial neutron and
final proton, $\gamma$ is a spin-orbit contribution, and $\beta$, $\delta$,
and $\varepsilon$ are three spin-spin terms.

The orthogonal unit vectors used in Eq.~\eqref{fpn} are defined in terms of
the initial neutron ($\vec{K}$) and final proton ($\vec{K'}$) cm momenta;
\begin{equation}
\vec{n}=\frac{\vec{K}\times\vec{K'}}{|\vec{K}\times\vec{K'}|},~~
\vec{m}=\frac{\vec{K'}-\vec{K}}{|\vec{K'}-\vec{K}|},~~
\vec{l}=\frac{\vec{K'}+\vec{K}}{|\vec{K'}+\vec{K}|}\,.
\end{equation}
The amplitudes are normalised such that the $np\to pn$
differential cross section has the form
\begin{equation}
\label{secpn} \left(\frac{\dd\sigma}{\dd t}\right)_{\!\!np\to pn}=
|\alpha(q)|^{2}+|\beta(q)|^{2}+2|\gamma(q)|^{2}
+|\delta(q)|^{2}+|\varepsilon(q)|^{2}\,.
\end{equation}

In the \Szero\ limit of very low $E_{pp}=k^2/m_p$, where $\vec{k}$ is the
$pp$ relative momentum, the deuteron charge exchange necessarily involves a
spin flip from the $S=1$ of the deuteron to the $S=0$ of the diproton. In
this case the contribution from the spin-independent amplitude $\alpha$ drops
out and one is left with only the spin-flip cross sections, as in the sum
rule of Eq.~\eqref{Rnp2}. The observables involving only the initial spins
that are accessible at ANKE are linked to the amplitudes
through~\cite{CAR1991,BAR1989}:
\begin{eqnarray}
\nonumber
\frac{d^{4}\sigma}{dtd^{3}k}&=&\fmn{1}{3}
I\left\{S^{-}(k,\half q)\right\}^2,\\
\nonumber
I\,A_y^d&=&0\:,\\
\nonumber
I\,A_y^p&=&-2\textit{Im}(\beta^*\gamma)\:,\\
\nonumber
I\,A_{xx}&=&{|\beta|^2+|\gamma|^2+|\varepsilon|^2-2|\delta|^2R^2}\:,\\
\nonumber
I\,A_{yy}&=&{|\delta|^2R^2+|\varepsilon|^2-2|\beta|^2-2|\gamma|^2}\:,\\
\nonumber
I\, C_{y,y}&=&-2\textit{Re}(\varepsilon^*\delta)R,\\
\nonumber
I\, C_{x,x}&=&-2\textit{Re}(\varepsilon^*\beta),\\
C_{yy,y} &=& -2A_y^p,
\label{impulse}
\end{eqnarray}
where the spin-flip intensity
\begin{equation}
\label{idce}
I=|\beta|^{2}+|\gamma|^{2}+|\varepsilon |^{2}+|\delta|^{2}R^{2}.
\end{equation}
The function
\begin{equation}
\label{ratio}
R=\left.{S^{+}(k,\half q)}\right/{S^{-}(k,\half q)}
\end{equation}
is the ratio of two transition form factors that involve the $S$- and
$D$-states of the deuteron wave function. In the forward direction $R=1$.

Since $\gamma(q)$ vanishes in the forward direction, the contributions of
$|\gamma(q)|^2$ to the cross section and the Cartesian tensor analysing
powers $A_{xx}$ and $A_{yy}$ in Eq.~\eqref{impulse} are almost negligible
under the conditions of the ANKE experiment. Measurements of the unpolarised
cross section and the two transverse analysing powers can therefore determine
separately the values of $|\beta(q)|^{2}$, $|\varepsilon(q)|^{2}$, and
$|\delta(q)|^{2}$ at fixed momentum transfer $q$. The two deuteron-proton
spin correlations that are measurable at ANKE, $C_{x,x}$ and $C_{y,y}$, fix
two of the relative phases. The proton analysing power $A_y^p$ gives mainly
information on the spin-orbit amplitude $\gamma(q)$.

Although the formulae given here describe the general features of our data,
detailed comparisons with theory are made using a program that takes higher
final $pp$ waves into account. These can, in particular, dilute the
polarisation signals~\cite{CAR1991}.

%
%
\section{The experimental facility}
\label{sec3}\setcounter{equation}{0}

The experiments reported here were carried out over three different time
periods using the ANKE magnetic spectrometer~\cite{BAR2001} that is placed at
an internal target position of the COoler SYnchrotron (COSY)~\cite{MAI1997}
of the Forschungs\-zentrum J\"ulich. Initially a polarised deuteron beam was
used in conjunction with an unpolarised hydrogen cluster
target~\cite{KHO1999}. In 2005 the \dpcepol\ reaction was studied at deuteron
beam energies $T_{d} = 1.2$, 1.6 and 1.8~GeV. The following year, the beam
energy was increased to 2.27~GeV, with 1.2~GeV being repeated for polarimetry
purposes. The equipment used was described in our earlier
publications~\cite{CHI2009,CHI2006a,CHI2006b}. However, for the study of the
spin-correlation parameters in 2009~\cite{KAC2007}, a newly developed
polarised internal target~\cite{GRI2007,MIK2012} was installed at ANKE and
this was employed in experiments at 1.2 and 2.27~GeV.

\subsection{The polarised deuteron beam at COSY}

The polarised deuterium ion source at COSY provides beams with different spin
configuration~\cite{CHI2006b}. It uses radio frequency transition units and
quadrupole magnets to exchange the occupation numbers of the hyperfine states
in the atom. The source was set up to provide a variety of states with
different tensor and vector polarisations but, as listed in
Table~\ref{pollist}, the selections differed for the experiments with cluster
or polarised cell target.

The COSY cycles were configured to provide beam first at 1.2~GeV and then,
without additional injection, accelerate the deuterons to one of the higher
energies. This procedure allows the use of the polarisation export
method~\cite{MCH2011}, which is crucial in the measurement of spin
observables at higher energies. This technique involves undertaking the
polarimetry measurements at the lowest $T_d = 1.2$~GeV flat top energy, where
the analysing powers are precisely known, and assuming that the beam
polarisation is unchanged at the higher energy. This procedure is viable
because there are no depolarising resonances for deuterons in the COSY energy
range. This was checked at the 4\% level by repeating the measurement of the
analysing powers after deceleration~\cite{MCH2011}.

\begin{table}[h]
\renewcommand{\arraystretch}{1.5}
\centering
 \begin{tabular}{c|c|c|c|c}
  Experiment        & State & $I_0$      & $P_{z}$          & $P_{zz}$ \\
  \hline
 unpolarised target &     1 & 1          & $\phantom{-}0$   & $\phantom{-}0$ \\
                    &     2 & 1          & $+\fmn{1}{3}$    & $+1$           \\
                    &     3 & 1          & $-\fmn{2}{3}$    & $\phantom{-}0$ \\
                    &     4 & 1          & $+\fmn{1}{3}$    & $-1$           \\
                    &     5 & 1          & $-\fmn{1}{3}$    & $+1$           \\
                    &     6 & \fmn{2}{3} & $\phantom{-}0$   & $+1$           \\
                    &     7 & \fmn{2}{3} & $\phantom{-}0$   & $-2$           \\
                    &     8 & \fmn{2}{3} & $-1$             & $+1$           \\
                    &     9 & \fmn{2}{3} & $+1$             & $+1$           \\
  \hline
 polarised target   &     1 & 1          & $\phantom{-}0$   & $\phantom{-}0$ \\
                    &     3 & 1          & $-\fmn{2}{3}$    & $\phantom{-}0$ \\
                    &     8 & \fmn{2}{3} & $-1$             & $+1$           \\
 \end{tabular}\vspace{1pc}
\caption{The different configurations of the polarised deuteron ion source
used in experiments carried out with the unpolarised cluster target, showing
the nominal (ideal) values of the vector ($P_{z}$) and tensor ($P_{zz}$)
polarisations and relative beam intensities. For the spin-correlation
measurements, the more restricted set was used. Note, that $P_{z}$ and
$P_{zz}$ are labeled conventionally in the reference frame of the
source~\cite{CHI2006b}.} \label{pollist}
\end{table}

\subsection{The polarised hydrogen cell target at ANKE}

The ANKE polarised internal gas target~\cite{MIK2012,ENG2008}
uses an Atomic Beam Source (ABS) that is capable of producing
both polarised hydrogen and deuterium beams~\cite{RAT2003}. The
first tests with the hydrogen atoms gave polarisations of
$+0.89 \pm 0.01$ and $-0.96 \pm 0.01$ for spin-up and
spin-down, respectively~\cite{MIK2012,MIK2011}.

The polarised atomic beam could be used directly from the source as a jet
target. During the commissioning runs, the ABS demonstrated an integral
jet-target thickness of about $1.5 \times 10^{11}~$cm$^{-2}$, which is
consistent with the predicted value~\cite{MIK2011}. However, much higher
target densities can be achieved  if one uses a storage cell fed by the ABS.
In order to achieve the maximum density, it is important to minimise the
dimensions of the storage-cell tube. But, on the other hand, this limits the
number of particles stored because of the beam heating and the consequent
losses on the cell walls. A maximum target density of about
$10^{13}~$cm$^{-2}$ was achieved with the cell during the commissioning runs
and this resulted in luminosities of up to $10^{29}~$cm$^{-2}$s$^{-1}$,
depending upon the beam intensity~\cite{ENG2008}.

During the 2009 beam time a cell made of $25~\mu$m thick aluminium foil
(99.95\% Al) was used~\cite{GRI2007}. In order to minimise depolarisation on
the cell surface, its inner walls were coated with Teflon. The cell had
dimensions $X\times Y\times Z=20\times15\times370~$mm$^3$, where $Z$ is
measured along the beam direction with $X$ and $Y$ referring to the
horizontal and vertical transverse directions, respectively.

A dedicated beam development was required to ensure that the COSY beam passed
successfully through the cell. Electron cooling~\cite{STE2003} and stacking
injection~\cite{KAM2004} were employed, with hundreds of injections per cycle
to increase the number of stored deuterons in the beam. In order to avoid
excessive background coming from the interactions of the beam halo particles
with the cell wall, scrapers were installed upstream of the target region.

In the hydrogen case, the ABS was configured to produce two
polarised states with equal gas densities. The polarisation of
the hydrogen target was flipped between spin-up ($\uparrow$)
and spin-down ($\downarrow$) every five seconds throughout the
whole COSY cycle, which lasted for one hour. Such a procedure
simplifies the later analyses by obviating the need to consider
the luminosities while calculating the asymmetries between
these states. In order to have a possibility to measure the
target spin-up $\uparrow$ and spin-down $\downarrow$
polarisations separately, runs with an unpolarised hydrogen
cell-target were also undertaken.
\begin{figure}[htb]
\begin{center}
\includegraphics[width=0.95\columnwidth]{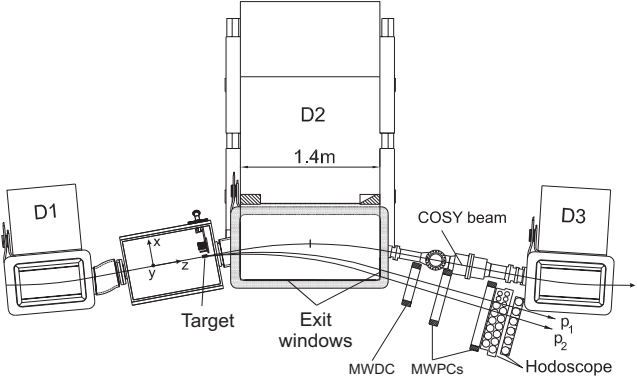}
\caption{\label{fig:ANKE} The ANKE experimental set-up showing
the positions of the three dipole magnets D1, D2, and D3. The
Forward Detector (FD) consists of multiwire drift (MWDC) and
multiwire proportional (MWPC) chambers, and a hodoscope
composed of three layers of scintillation counters. For
measurements with the polarised target, the ABS replaced the
cluster-jet in the target chamber, with the detection system
being identical in the two cases. The axes of the coordinate
system are indicated.}
\end{center}
\end{figure}

\subsection{The ANKE detection system}

\begin{figure}[htb]
\begin{center}
\includegraphics[width=0.9\columnwidth]{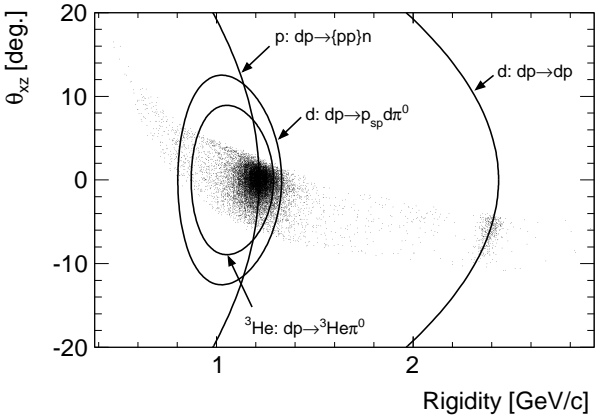}
\caption{\label{fig:rigidity} Scatter plot of singly charged particles
detected in ANKE from the interaction of 1.2~GeV deuterons with a hydrogen
cluster-jet target in terms of the laboratory production angle in the
horizontal plane and the magnetic rigidity. The loci corresponding to four
common nuclear reactions are also shown. These include the \dpce\ reaction at
zero $E_{pp}$.}
\end{center}
\end{figure}

The forward detector system (FD) of the ANKE magnetic spectrometer,
illustrated in Fig.~\ref{fig:ANKE}, is used for the deuteron charge-exchange
studies. The multiwire chambers in the FD serve for track reconstruction and
the three layers of the scintillation hodoscope permit the measurements of
the arrival time and energy-loss that are required for particle
identification~\cite{DYM2004}. The main trigger used in the experiments
consisted of a coincidence between the different layers in the hodoscope of
the FD. Figure~\ref{fig:rigidity} shows the experimental yield of ANKE for
single charged particles at $T_d=1.2$~GeV in terms of the laboratory
production angle in the horizontal plane and the magnetic rigidity. The
kinematic curves for some of the possible nuclear reactions are also
illustrated.

\begin{figure}[h]
\centering
\includegraphics[width=0.9\columnwidth]{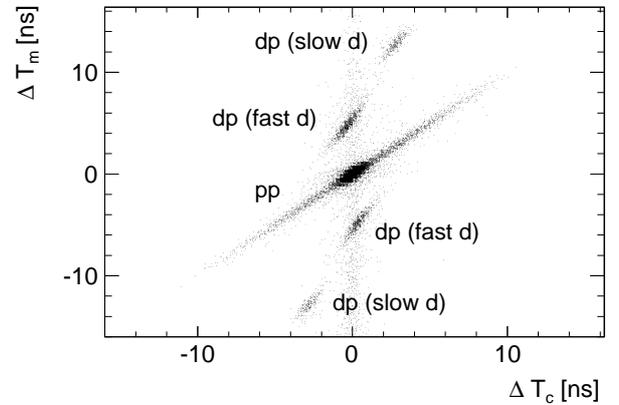}
\caption{Scatter plot of measured ($\Delta T_m$) and calculated ($\Delta
T_c$) time differences between pairs of charged particle registered in
the ANKE forward detector at $T_d=1.2$~GeV. $\Delta T_c$ was calculated
assuming that both particles were protons. The separation between $dp$ and
$pp$ pairs is very clear.} \label{fig:tctm}
\end{figure}

Among the reactions observed, there are two that are of
particular interest, namely the deuteron charge-exchange \dpce\
and the quasi-free \dpdppi, where the proton, $p_{\rm sp}$, has
about half the beam momentum. The latter reaction is used to
measure both the vector polarisation of the deuteron beam or
hydrogen target and also the luminosity. After recording two
charged particles, deuteron-proton pairs are separated from the
remaining two-track events (mainly proton pairs) in the
subsequent analysis by using the time information from the
hodoscope. As demonstrated in Fig.~\ref{fig:tctm}, if one
assumes that both detected particles in the pair are protons,
the calculated ($\Delta T_c$) time of flight difference from
the target does not match with the measured one ($\Delta T_m$)
for other pairs. After recognising the two charged particles,
the missing-mass distribution allows one to identify the
reaction.

\section{The cross section determination}
\label{sec4}\setcounter{equation}{0}

The cross section $\sigma$ for a given physical process is given in terms of
the corresponding counting rate $R$ and the luminosity $L$ through:
\begin{equation}
  \sigma = {R}/{L}
\label{lum}
\end{equation}
The luminosity, which is the product of the target density and beam
intensity, can be measured in various ways. In the current analysis we relied
on the measurement in parallel of the rate for a process with a well-known
and sizeable cross section. Once the luminosity is known, absolute values of
cross sections for other reactions can be deduced from the count rates
measured in the experiment.

\subsection{The $\boldsymbol{dp \rightarrow p_{\rm sp}d\pi^{0}}$ reaction}
\label{dpizero}

The \dpdppi\ reaction is used in the determination of the
luminosity in this experiment. This is identified in the ANKE
forward detector by detecting both charged particles (cf.\
Figs.~\ref{fig:rigidity} and \ref{fig:tctm}). After recognising
the $dp$-pairs, the reaction is finally isolated on the basis
of the missing-mass distributions~\cite{CHI2009}. There is an
accidental background at very small $|\Delta T_c|$ that is
randomly distributed in $\Delta T_{m}$. This is caused by fast
particles, mainly protons, that are produced in a different
beam-target interaction. The contribution from such accidental
events in the vicinity of the fast deuteron branch of the
\dpdppi\ reaction increases rapidly with energy. It varies
between $18$\% and $30$\% for $T_{d}\geq 1.6$~GeV, whereas it
is less then $3$\% at 1.2~GeV. The background is negligible at
all energies for the slow deuteron branch.

The properties of background were studied using the data for which $|\Delta
T_c|< 2$~ns and $|\Delta T_m|> 12$~ns. As can be seen from
Fig.~\ref{fig:tctm}, no true coincidence two-track events are expected in
this region. These data provided the shape of background in the distributions
of the missing mass and the deuteron laboratory scattering angle in the
\dpdppi\ reaction. The normalisation for the background was found by
comparing the background missing-mass spectra with that for the identified
$dp$ pairs. The normalised background was then subtracted from the \dpdppi\
angular distributions.

\begin{figure}[htb]
\centering
\includegraphics[width=0.8\columnwidth]{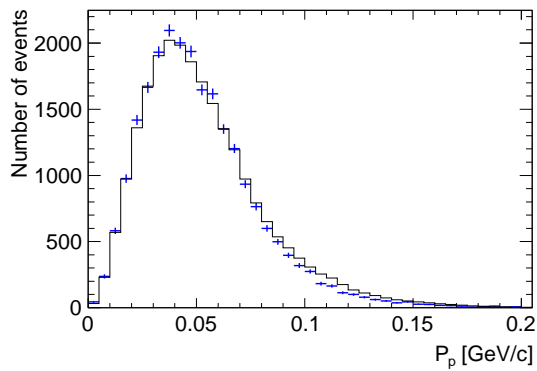}
\caption{The momentum distribution of the fast proton from the \dpdppi\
reaction at $T_{d}=1.2$~GeV, transformed into the rest frame of the incident
deuteron is compared with the Monte Carlo simulation (solid histogram).}
\label{fig:Pf}
\end{figure}

At intermediate energies, soft deuteron collisions are
generally dominated by the interaction of one of the nucleons
in the nucleus, the other nucleon being  a spectator. When the
proton acts as a spectator, $p_{\rm sp}$, the \dpdppi\ reaction
can be interpreted in terms of quasi-free \npdpi\ pion
production. To confirm the spectator hypothesis, a Monte Carlo
simulation has been performed within PLUTO~\cite{PLUTO} using
the Fermi momentum distribution from the Paris deuteron wave
function~\cite{LAC1981}. As is clear from Fig.~\ref{fig:Pf},
the data are consistent with quasi-free production on the
neutron leading to a spectator proton. However, in order to
reduce further possible contributions from multiple scattering
and other mechanisms, only events below 60~MeV/$c$ were
retained for the luminosity evaluation.

The determination of the angles for the quasi-free \npdpi\
reaction is complicated by the Fermi motion of the nucleons
inside the deuteron. Due to this effect, the effective neutron
beam energy, $T_n$, is spread around half the deuteron beam
energy with a width arising from the Fermi momentum. At a beam
energy of $600$~MeV per nucleon, the FWHM is $90$~MeV for a
$p_{\rm{}sp}<60$~MeV/$c$ cut. Furthermore, the neutron
direction is not precisely aligned along that of the beam, but
is spread over some solid angle. Since this introduces an
incident angle, which is several degrees in the laboratory
system (depending on the beam energy), it has to be taken into
account. These considerations apply to both the polar and
azimuthal angles. In order to correct for this effect, the
three-momentum of the incident neutron was reconstructed using
the information from the spectator-proton momentum. The
deuteron polar angle was measured from the neutron momentum
instead of the beam direction. The azimuthal angle was defined
between the normals to the COSY ring and deuteron scattering
plane.

Isospin invariance requires the cross section for $np \to d\pi^0$ to be half
of that for \ppdpi, for which there are numerous measurements~\cite{ARN1993}.
An additional advantage of using this reaction for normalisation is that the
typical 5\% shadowing effect in the deuteron (where one nucleon hides behind
the other) should be broadly similar in the $dp\to \{pp\}X$ and \dpdppi\
reactions.

\begin{figure}[htb]
 \begin{minipage}{10pc}
  \centering
  \includegraphics[width=1\columnwidth]{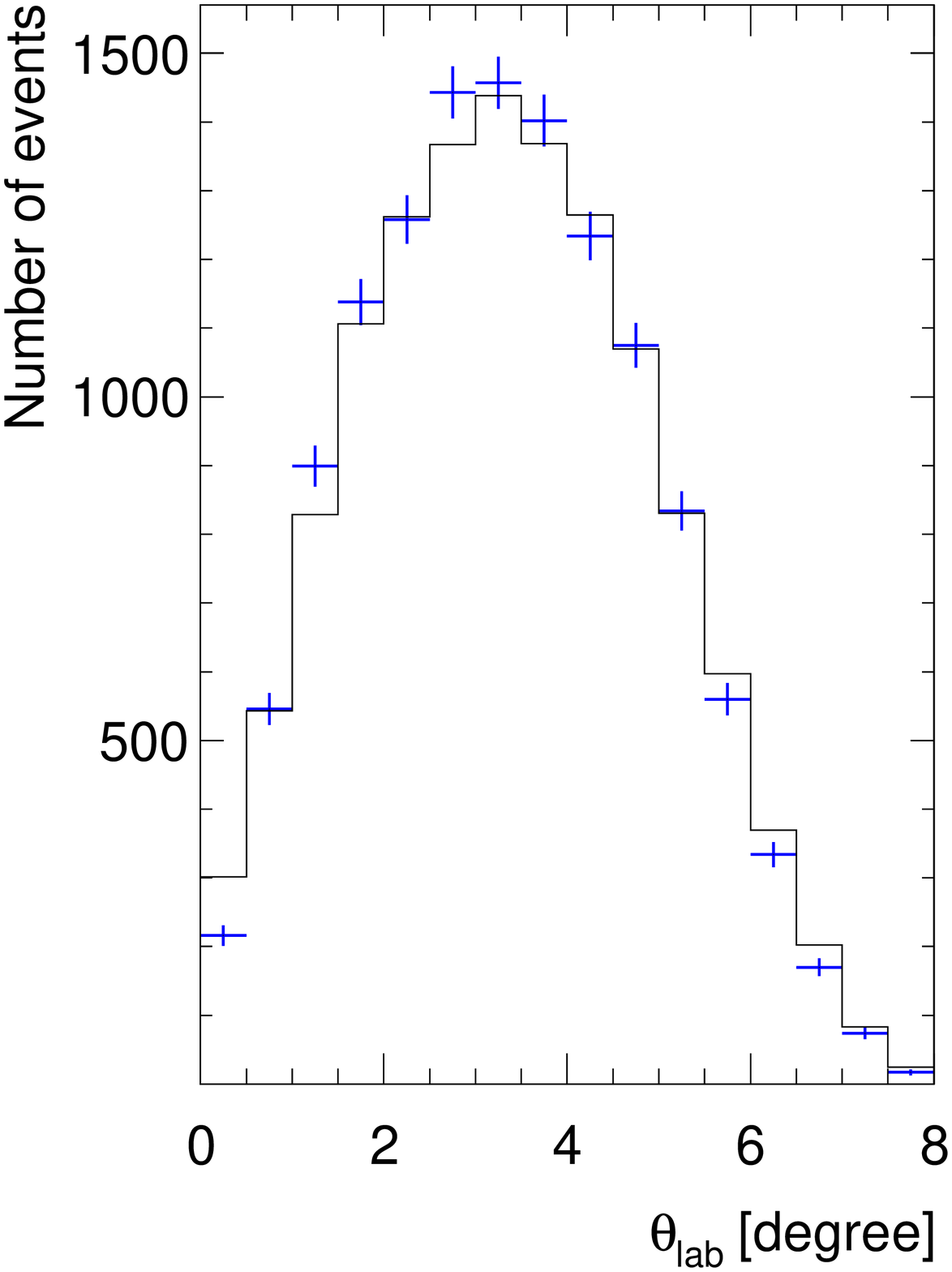}
 \end{minipage} \vspace{2pc}
 \begin{minipage}{10pc}
  \centering
  \includegraphics[width=1\columnwidth]{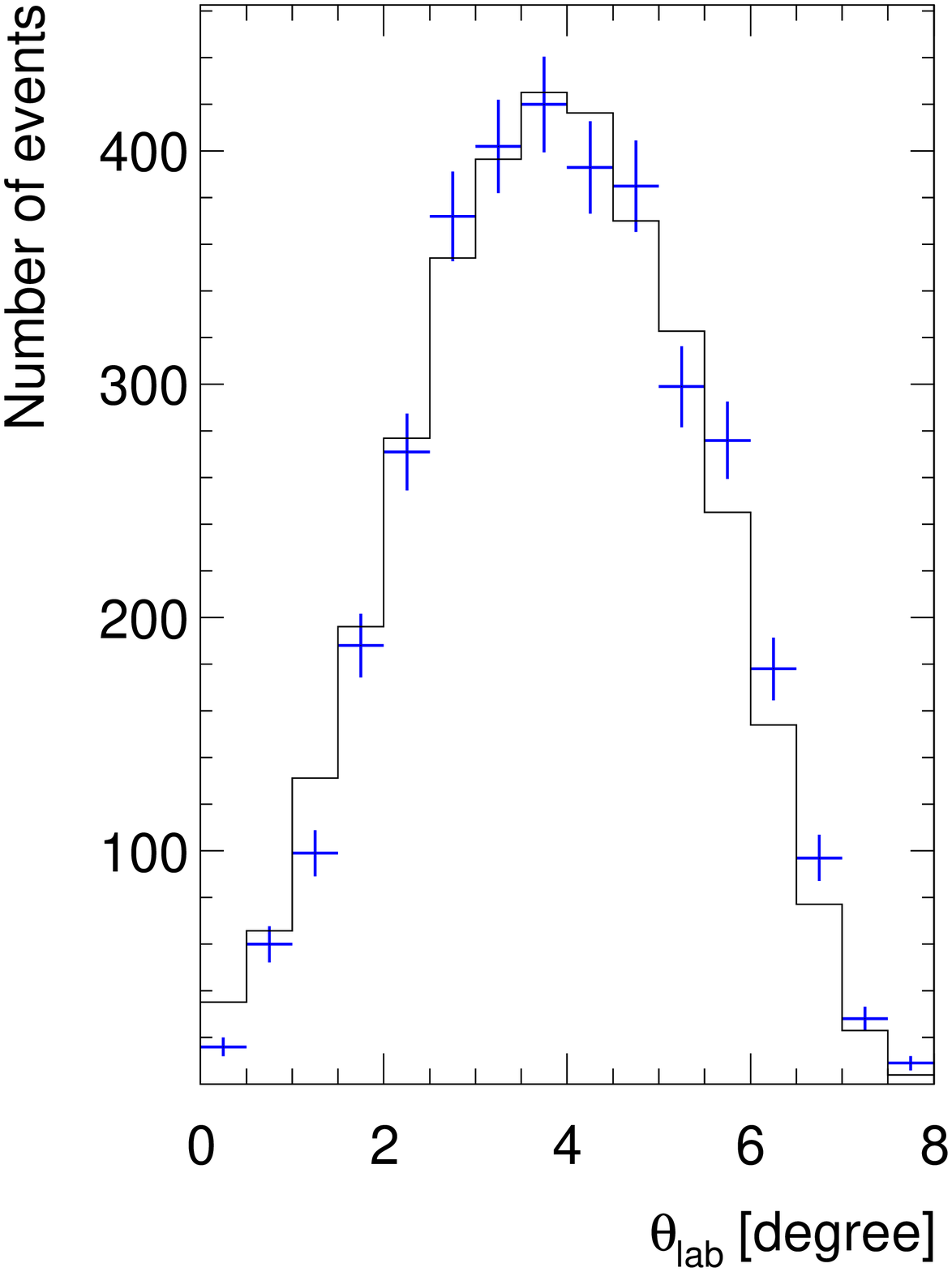}
 \end{minipage}
\caption{Simulated (solid histogram) and experimental angular distributions
for the \dpdppi\ reaction at $T_{d}=1.2$~GeV. The left and
right panels correspond to fast and slow deuterons, respectively.} \label{Dangle}
\end{figure}

In order to investigate the acceptance of the ANKE forward detector for
different reactions, a full simulation was performed based on GEANT
software~\cite{AGO2003}. The same track reconstruction algorithm was used in
the simulation and the data analysis. In order to get as precise a
description of the experiment as possible, the dispersion of the hits in the
MWPC, the background hits produced by accidental coincidences, and the noise
in the multiwire chambers readout electronics, as obtained from the
experimental data, were also included in the simulation~\cite{DYM2004}. The
quality of the simulation may be judged from the distributions of the
deuteron production angle in the laboratory system that are shown in
Fig.~\ref{Dangle}.

\subsection{Luminosity measurements}

To determine the luminosity we use Eq.~(\ref{lum}) and insert detailed
expressions for $R$ and $\sigma$:
\begin{equation}
 L = \frac{R}{\sigma} = \frac{ R_{\rm exp}\, ({N_{\rm tot}}/{N_{\rm acc}}) }
  { \iint\frac{d\sigma}{d\vartheta}(\vartheta,T_{n})\,d\vartheta\, dT_{n} }\,,
 \label{Lum}
\end{equation}
where $R_{\rm exp}$ is the count rate from the quasi-free $np\to d\pi^0$
reaction, corrected for the trigger dead time. $N_{\rm acc}$ is the number of
counts in the simulation that pass all the criteria used in the experimental
data processing and $N_{\rm tot}$ is the total number of simulated events.
These are summed over the neutron kinetic energy, subject to the $p_{\rm
sp}<60$~MeV/$c$ cut, and over the given angular range. The
${d\sigma}(\vartheta,T_{n})/{d\vartheta}$ differential cross section was
taken from the $pp\to d\pi^+$ database~\cite{ARN1993}, where $\vartheta$ is
the deuteron polar angle in the laboratory frame.

The FD detector acceptance changes rapidly with angle. In order to minimise
systematic errors, the total angular range was binned and the luminosity
evaluated separately for each bin. Data at the acceptance edges (the smallest
and the largest angles) were less reliable, due to the greater uncertainty in
the evaluation of the acceptance, and showed systematic shifts in luminosity.
Such angular intervals were discarded and the average recomputed.

The values of the average luminosities determined from the
$np\to d\pi^0$ reaction are given in Table~\ref{avlum}. The
errors quoted include statistical ones from the experimental
counts and those introduced by the background subtraction
procedure. Uncertainties coming from the SAID
database~\cite{ARN1993}, which were estimated by studying the
experimental results in the relevant regions, are listed
separately.

\begin{table}[htb]
\renewcommand{\arraystretch}{1.5}
\begin{center}
\begin{tabular}{c|c|c|c}
 $T_{d}$          & Average luminosity   & Measurement & SAID\\
 {}[GeV]          & [cm$^{-2}$s$^{-1}$]  & uncertainty & uncertainty\\
                  &                      & [\%]        & [\%]\\
 \hline
 $1.2\phantom{1}$ & $1.76\times 10^{30}$ & $1.1$       & $2.2$\\
 $1.6\phantom{1}$ & $1.84\times 10^{31}$ & $2.0$       & $5.1$\\
 $1.8\phantom{1}$ & $1.61\times 10^{31}$ & $2.8$       & $4.4$\\
 $2.27$           & $1.18\times 10^{30}$ & $5.0$       & $3.8$\\
\end{tabular}
\vspace{2mm} \caption{Average luminosities achieved with the
cluster-jet target at four different beam energies. Shown
separately are the uncertainties associated with the
measurement and with the experimental data used as input in the
estimations.} \label{avlum}
\end{center}
\end{table}

\subsection{The $\boldsymbol{dp\to \{pp\}_{\!s}n}$ cross section}

\begin{figure}[htb]
\begin{center}
\includegraphics[width=0.8\columnwidth]{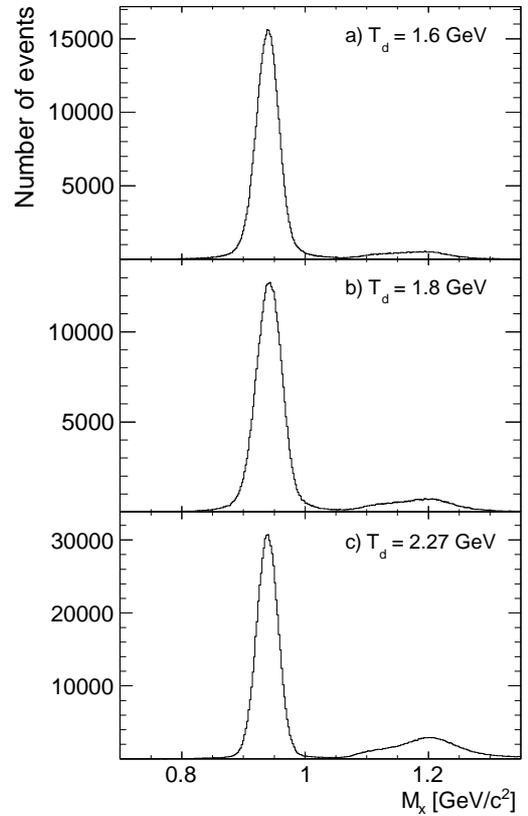}
\caption{\label{MM_pp} The missing-mass $M_x$ distributions for the $dp\to
\{pp\}_{\!s}X$ reaction at three deuteron beam energies. The background
under the neutron peak is a negligible.}
\end{center}
\end{figure}

Having identified two fast protons in the final state and
selected low $E_{pp}$ events, the \dpce\ reaction was isolated
on the basis of the missing-mass distributions, which are shown
at three energies in Fig.~\ref{MM_pp}. In addition to the
dominant neutron peak, there are also many events for
$M_x>1080$~MeV/$c^2$ that must correspond to pion
production~\cite{MCH2011a}. However, very few of these leak
into the neutron region and the background from this under the
neutron peak is at most at the per cent level. This is also
true for data in the individual momentum-transfer $q$ bins.
Random background, which was studied using the timing
information, was at the 1-3\% level and could be easily
subtracted.

The \dpce\ cross section determination was performed in a
similar manner to the luminosity evaluation that used the
\dpdppi\ reaction. It involved the same technique for
correcting the experimental count rates and estimating the
detector acceptance. Since the \dpce\ reaction has a three-body
final state, in principle the cross section is a function of
five independent variables. However, within the impulse
approximation, by far the most important of these are the
excitation energy $E_{pp}$ in the final $pp$ diproton and the
momentum transfer $q$ from the proton to the neutron. In
accordance with Eq.~(\ref{lum}), the two-dimensional
differential cross section was evaluated in terms of the
integrated luminosity $L_{\rm int}$ from:
\begin{equation}
 {\frac{d^2\sigma(q,E_{pp})}{dq\,dE_{pp}}} = \frac{1}{L_{\rm int}}\,
  \frac{N_{\rm exp}(q,E_{pp})\, N_{\rm tot}(q,E_{pp})}{N_{\rm acc}(q,E_{pp})
  \, \Delta q \, \Delta E_{pp}}\,,
 \label{CS}
\end{equation}
where $N_{\rm exp}(q,E_{pp})$ is the corrected number of
experimental events for given values of $q$, measured in
laboratory system, and $E_{pp}$. $N_{\rm tot}(q,E_{pp})$ and
$N_{\rm acc}(q,E_{pp})$ are the total and accepted numbers of
simulated events respectively. $\Delta q$ and $\Delta E_{pp}$
correspond to bin widths in momentum transfer and excitation
energy, respectively. The resolutions in $q$ and $E_{pp}$ are
about $4-8$~MeV/$c$ and better than $0.3$~MeV, respectively.

\begin{figure}[htb]
\begin{center}
\includegraphics[width=0.8\columnwidth]{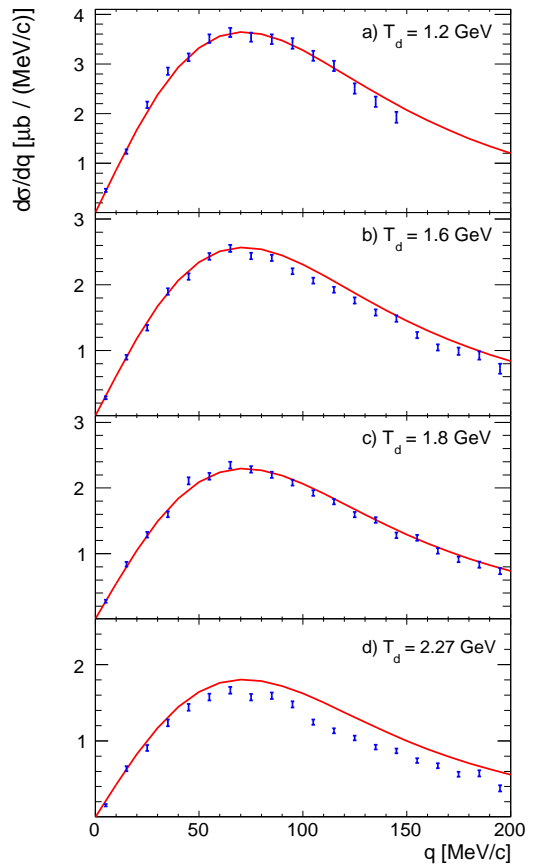}
\caption{\label{fig_cs} Differential cross sections for the \dpce\ reaction
at four different energies compared with impulse approximation predictions
based upon the current SAID $np \to np$ amplitude analysis~\cite{ARN2000}.
The data are integrated over the $E_{pp}<3$~MeV interval. Only statistical
errors are shown.}
\end{center}
\end{figure}

The cross sections were further integrated over $E_{pp}<3$~MeV in order to
provide the $d\sigma/dq$ differential distribution presented in
Fig.~\ref{fig_cs}. This includes also the new results obtained at
$T_d=1.2$~GeV. In addition to the statistical errors arising from the
experimental count rates that are shown, there are also overall systematic
uncertainties arising from the luminosity determinations, given in the
Table~\ref{avlum}. Within these uncertainties, the agreement with the
theoretical impulse approximation predictions~\cite{CAR1991} at $T_d = 1.2$,
1.6, and $1.8$~GeV is very encouraging and is in line with similar data
analysed at 1.17~GeV~\cite{CHI2009}. In contrast, the unpolarised
differential cross section at $T_d = 2.27$~GeV falls about 15\% below the
predictions based upon the current $np \to np$ partial wave
analysis~\cite{ARN2000}. As we shall see later, similar discrepancies are
found in the spin observables of the \dpce\ reaction but only at this highest
energy.

\section{Determination of deuteron analysing powers}%
\label{sec5}\setcounter{equation}{0}

In order to measure the deuteron analysing powers for the \dpcepol\ or other
reaction, the first step has to be the identification of the polarisations of
the various deuteron beams used in the experiment. With complete efficiencies
in the transition units, the polarisations should approach the ideal values
given in Table~\ref{pollist}. However, this is never the case in practice and
the beam polarisation must be determined separately for each of the beams
used. In our earlier work at 1.17~GeV~\cite{CHI2009}, a variety of nuclear
reactions with known analysing powers were measured and these were used to
establish values for the polarisations. These showed that the analysing
powers for the \dpcepol\ reaction were well reproduced in the impulse
approximation calculations~\cite{CAR1991}. Since the deuteron charge-exchange
reaction can be so well identified and measured at ANKE, we use this reaction
itself to measure the beam tensor polarisation at the neighbouring energy of
1.2~GeV, the necessary analysing powers being taken from the impulse
approximation estimates. Apart from the large counting rates, this approach
has the advantage of being insensitive to the deuteron vector polarisation.

\subsection{Tensor polarisation of the deuteron beam}

The tensor polarisation of the deuteron beam was measured using the \dpce\
reaction at $T_d = 1.2$~GeV. The circulating deuteron beam was polarised
perpendicularly to the horizontal plane of the machine. As already mentioned,
the beam vector ($P_{z}$) and tensor ($P_{zz}$) polarisations are labeled
conventionally in the reference frame of the source. In contrast, all the
spin observables discussed later refer to the right-handed coordinate system
of the reaction frame, where the beam defines the $Z$-direction while the
stable spin axis of the beam points along the $Y$-direction, which is
perpendicular to the COSY orbit.

No dependence is expected on the vector polarisation for small
$E_{pp}$~\cite{BUG1987} and this was checked in our earlier
experiment~\cite{CHI2009}. The numbers $N(q,\phi)$ of diprotons produced at
momentum transfer $q$ and azimuthal angle $\phi$ with respect to the
$X$-direction are given in terms of the beam polarisation by
\begin{eqnarray}
\nonumber \frac{N(q,\phi)}{N_{0}(q)} = C_n\left\{1+\fmn{1}{4}P_{zz}
\left[A_{xx}(q)(1-\cos2\phi) + \right. \right. \\
\left.\left.A_{yy}(q)(1+\cos2\phi)\right]\right\},
\label{eq:ce_1}
\end{eqnarray}
where $N_{0}(q)$ are the numbers for an unpolarised beam and
$C_n$ is the relative luminosity of the polarised beam.

During the course of the polarised measurements, various configurations of
the ion source were used and the beam polarisation had to be determined
separately for each state. Several methods to fix the relative luminosities
$C_n$ of a state with respect to the unpolarised mode are possible at
ANKE~\cite{MCH2011}. The one provided by the $dp \rightarrow p_{\rm sp}X$
reaction is preferable because the number of single-track events is enormous
for all beam energies. No dependence of the rates on the tensor polarisation
of the deuteron beam was found for proton spectator momenta below
$60$~MeV/$c$.

The break-up data were divided into several bins of momentum transfer $q$ and
distributions in $\cos2\phi$ were constructed for each bin and polarisation
mode. The ratios to the unpolarised state were fitted using
Eq.~(\ref{eq:ce_1}), the theoretical predictions for $A_{xx}$ and $A_{yy}$
being taken at mean values of $q$ in each bin. The validity of this approach
was checked at $T_d = 1.17$~GeV in the earlier experimental studies at
ANKE~\cite{CHI2006a}. The beam polarisation in each state was taken as the
weighted average over the different values of the momentum transfer.

The maximum values of $P_{zz}$ were $\approx85\%$ of the ideal
values for the 2005 data. But, in the 2006 data, the maximum
tensor polarisation dropped to $\approx 55\%$ of the ideal,
with little change in the vector polarisation. The difference
has been ascribed to the changed efficiencies of the units in
the COSY deuterium ion source~\cite{MCH2011}. For the high
$|P_{zz}|$ modes in Table~\ref{pollist}, the estimated
polarisation varied in $(0.75-0.85)$ range in the 2005 data and
in the $(0.39-0.60)$ range in the 2006 data. The typical
uncertainty in $P_{zz}$ was $0.02-0.04$.

The vector analysing power of the \dpcepol\ reaction is predicted to vanish
in the \Szero\ limit~\cite{BUG1987} and this was confirmed at 1.17~GeV in our
earlier work~\cite{CHI2009}. As a consequence, the vector polarisation of the
beam is unimportant for the tensor analysing power studies carried out with
the cluster target. This is no longer the case for the spin-correlation
measurements with the polarised cell target. The determination of the vector
polarisation of the deuteron beam in this case is described in
sec.~\ref{sec6}.

\begin{figure}[ht]
\begin{center}
\includegraphics[width=0.8\columnwidth]{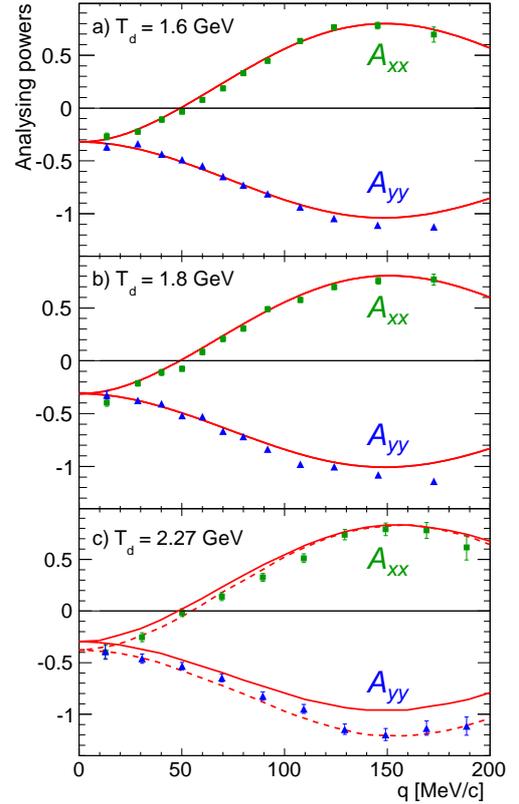}
\caption{\label{fig:AP}Tensor analysing powers $A_{xx}$
(squares) and $A_{yy}$ (triangles) of the \dpcepol\ reaction at
three beam energies for low diproton excitation energy,
$E_{pp}<3$~MeV, compared to impulse approximation predictions
based upon the current SAID $np\to np$
amplitudes~\cite{ARN2000}. The dashed curves at $2.27$~GeV
correspond to a uniform reduction of the spin-longitudinal
$\varepsilon(q)$ amplitude by $25\%$. The error bars include
the uncertainties from the beam polarisation and relative
luminosity $C_{n}$.}
\end{center}
\end{figure}

\subsection{Deuteron tensor analysing powers}

The deuteron Cartesian tensor analysing powers $A_{xx}$ and $A_{yy}$ were
extracted using Eq.~(\ref{eq:ce_1}) in much the same way as for the
polarimetry, with the beam tensor polarisation $P_{zz}$ being determined at
$T_d = 1.2$~GeV. The ratios of the polarised to unpolarised corrected count
rates were fitted in terms of the two free parameters $A_{xx}$ and $A_{yy}$.
The procedure was repeated for different polarisation states and results
averaged over these source modes.

The results for the tensor analysing powers are shown in Fig.~\ref{fig:AP} at
three beam energies as functions of the momentum transfer. The agreement
between the experimental data and the impulse approximation predictions is
very good at $T_n = 800$ and 900~MeV. At these energies the SAID $np$
amplitudes~\cite{ARN2000} used as input in the calculations are considered to
be reliable but at 2.27~GeV, the maximum deuteron energy available at COSY,
the agreement is much worse. Since there are also severe discrepancies in the
unpolarised cross section at this energy, it is natural to question whether
there might be deficiencies in the SAID $np$ analysis at this energy.

The experimental value of $A_{xx} = A_{yy}$ in the forward
direction ($q\approx0$) is significantly more negative than the
predictions using the SAID $np$ amplitudes at 1.135~GeV.
However, it can be seen from Eq.~(\ref{impulse}) that a
relative reduction in the $\varepsilon$ amplitude might improve
the predictions. To check this possibility, the predictions
were recomputed with the $\varepsilon(q)$ amplitude being
reduced uniformly by 25\%. This gives the much better overall
agreement with the data that is demonstrated by the dashed
curves in the lower panel of Fig.~\ref{fig:AP}. This therefore
suggests that the current SAID amplitudes~\cite{ARN2000} might
overestimate the relative strength of the $\varepsilon(q)$ at
small $q$ but further proof is required and this is furnished
by the measurements of the spin correlations.

\section{Determination of the deuteron-proton spin-correlation parameters}
\label{sec6}\setcounter{equation}{0}

In experiments with the unpolarised cluster-jet target, the \dpcepol\
reaction is only sensitive to the tensor polarisation of the beam and the
values of this could be established by using the \dpcepol\ reaction itself at
the 1.2~GeV calibration energy, as described in sec.~\ref{sec5}. In contrast,
in order to determine the spin-correlation parameters $C_{x,x}$ and
$C_{y,y}$, one has first to determine the vector polarisation of the deuteron
beam as well as the polarisation of the hydrogen in the target cell. The
basis of both measurements is the quasi-free $\pol{n}p\to d\pi^0$ reaction
which, at small spectator momenta, is insensitive to the deuteron tensor
polarisation.

\subsection{The beam and target polarimetry}

The polarisation of the hydrogen target and the vector polarisation of the
deuteron beam were both determined using the quasi-free \npdpi\ reaction.
Now it is well known that, if one integrates over all Fermi momenta inside
the deuteron, the nucleon polarisation in the deuteron $P_z^n$ would be
reduced from that of the deuteron $P_z^d$ by a factor
\begin {equation}
P_z^n = \left(1-\fmn{3}{2}P_D\right) P_z^d\,, \label{eq:Pol_n}
\end {equation}
where $P_D$ is the deuteron $D$-state probability. However, since the
$D$-state effects vanish like $(p_{\rm sp})^2$, the dilution of the
polarisation signal by the deuteron $D$-state is negligible if only data with
$p_{\rm sp}<60$~MeV/$c$ are used in the subsequent analyses. Such a cut
preserves a large part of the statistics.

For an unpolarised deuteron beam incident on a polarised hydrogen target with
spin-up ($\uparrow)$ and spin-down ($\downarrow$), the asymmetry ratio
$\epsilon$ between polarised $N^{\uparrow (\downarrow)}$
and unpolarised $N^{0}$ yields has the form:
\begin {equation}
\epsilon^{\uparrow (\downarrow)}(\theta, \phi) = \frac{N^{\uparrow
(\downarrow)}(\theta,\phi)}{N^{0}(\theta)} = 1 + Q^{\uparrow (\downarrow)}
A_{y}(\theta) \cos\phi, \label{eq:dc_asymm}
\end {equation}
where $\theta$ and $\phi$ angles are polar and azimuthal angles,
respectively, of the detected particle and $Q$ is the target polarisation.
Isospin invariance requires the analysing powers in the \npdpi\ and \ppdpi\
reactions to be identical and there are numerous measurements of the proton
analysing power $A_{y}(\theta)$ of the latter in the 600~MeV
region~\cite{ARN1993}. Data in the centre-of-mass system were used in the
analysis. The polar $\vartheta$ and azimuthal $\phi$ angles of the deuteron
were determined according to the procedure described in sec.~\ref{dpizero}.

\subsection{The $\boldsymbol{dp\to p_{\!sp}d\pi^{0}}$ reaction with the cell target}

The cell introduces additional complications in the determination of the
angles because of the spread of the interaction points along the cell axis.
The reconstruction of the longitudinal vertex coordinate $Z$ is therefore
required for each event. For a two-track event, this can be done with the use
of the arrival-time difference for two particles, measured in the
scintillation hodoscope. In our kinematical conditions, where the deuteron is
at least twice as slow as the proton, such a difference is a sensitive
function of $Z$. The three-momenta of the two particles and the $Y$ and $Z$
coordinates of the vertex are found through an overall fit procedure that
uses the information from both the wire chambers and the hodoscope.

\begin{figure}[ht]
\begin{center}
\includegraphics[width=0.9\columnwidth]{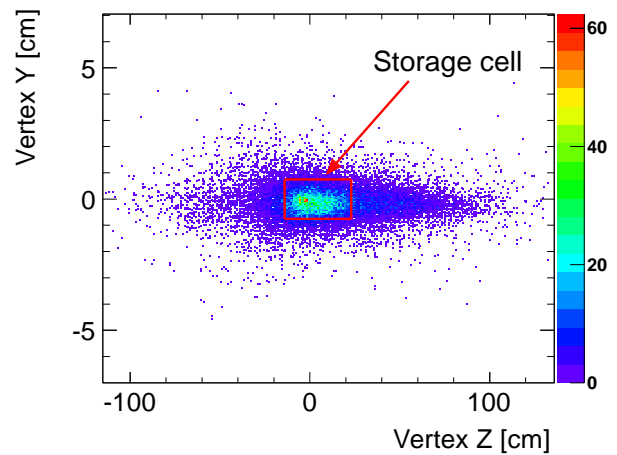}
\caption{\label{fig:dc_vertex} Vertex reconstruction in the $YZ$ plane using
correlated deuteron-proton pairs. The rectangle shows the physical dimensions
of the cell ($Y\times Z = 15\times 370$ mm$^{2}$).}
\end{center}
\end{figure}

Figure~\ref{fig:dc_vertex} shows the distribution of interaction points in
the $Y$-$Z$ plane. In addition to helping in the angular determination, the
vertex reconstruction allows one to make cuts along the cell axis to minimise
the background from the rest gas that is spread throughout the target
chamber.

\begin{figure}
\begin{center}
\includegraphics[width=0.8\columnwidth]{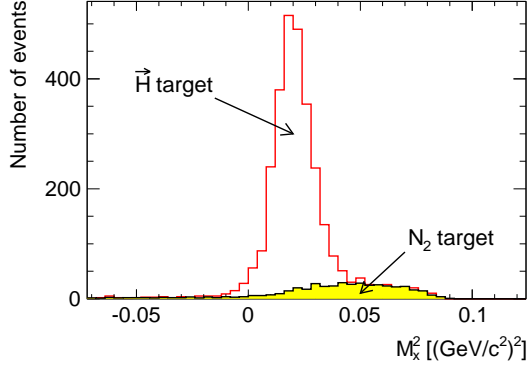}
\caption{\label{fig:dc_bgsub} Comparison of the $(d,dp_{\rm sp})$
missing-mass-squared distributions at $T_{d}=1.2$~GeV when using a polarised
hydrogen target or filling the cell with nitrogen gas.}
\end{center}
\end{figure}

A second major complication arises from the scattering of the beam halo
particles on the cell walls. This can produce additional background that
would dilute the analysing power signal. As mentioned earlier, the dedicated
beam development enabled the bulk of the beam to pass through the cell
without hitting the walls. For this reason, recording data with an empty cell
would take much more time to collect sufficient statistics to determine the
background. Additional runs were therefore recorded where nitrogen gas was
injected into the cell to simulate the shape of the background (see details
in Ref.~\cite{GRI2007}). The background subtraction was performed for each
polarisation state by using the missing-mass distributions for the hydrogen
and nitrogen data. One such example is shown in Fig.~\ref{fig:dc_bgsub}.

\subsection{The target and beam polarisation}

The target polarisation was measured using data taken with an unpolarised
deuteron beam at $T_{d}=1.2$~GeV. After vertex reconstruction, the \npdpi\
data were binned in deuteron cm angles. The background subtraction was
performed separately for each bin and  distributions in $\cos\phi$ built for
both spin-up and spin-down modes. The weighted sum of the two data sets from
different target polarisations was taken as the unpolarised mode. Weights
were determined according to the relative asymmetries with respect to the
unpolarised state. The ratios of the difference to the sum of the data for
the two polarised modes were than fitted with a linear function in $\cos\phi$
and the value of the product $Q A_{y}$ deduced. Taking the mean analysing
power $\langle A_y \rangle$ in each $\theta_{d}^{cm}$ bin from the SAID
\ppdpi\ database~\cite{ARN1993}, this gave $Q^{\uparrow} = 0.61 \pm 0.02$ and
$Q^{\downarrow} = -0.70 \pm 0.03$.

Using the quasi-free \npdpi\ reaction, in an analogous way to that for the
target polarisation, the vector polarisation of the deuteron beam (state 3 in
Table~\ref{pollist}), was determined to be $P_z=-0.51\pm 0.05$.

\subsection{Measurement of the deuteron-proton spin-correlation parameters}%

The double-polarised experimental data from the 2009 beam time offers an
excellent opportunity for studies of the spin-correlation parameters
$C_{x,x}$ and $C_{y,y}$ to determine the relative phases of the spin-spin
amplitudes~\cite{KAC2007}. Furthermore, it is of interest to see whether the
suspected $\varepsilon$ deficiencies in the SAID amplitudes at highest beam
energy are reflected also in the spin correlations.

\begin{figure}[htb]
\begin{center}
\includegraphics[width=0.8\columnwidth]{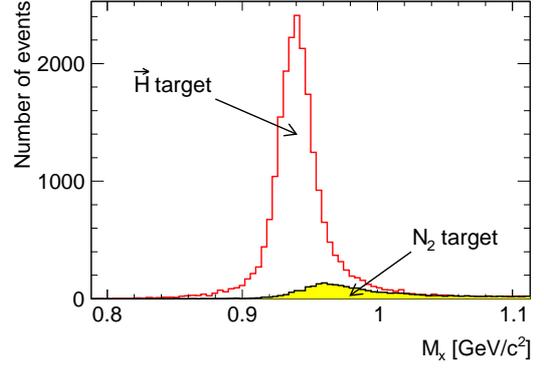}
\caption{\label{fig:dc_bgppn} Comparison of the $(d,pp)$ missing-mass
distributions at $T_{d}=1.2$~GeV beam energy when using a polarised hydrogen
target or filling the cell with nitrogen gas.}
\end{center}
\end{figure}

The background subtraction was carried out in the same way as for the
$dp\rightarrow d\pi^{0}p_{\rm sp}$ reaction, with the nitrogen gas data
simulating the shape of the background, as illustrated in
Fig.~\ref{fig:dc_bgppn}.

\begin{figure}[h]
\begin{center}
\includegraphics[width=0.8\columnwidth]{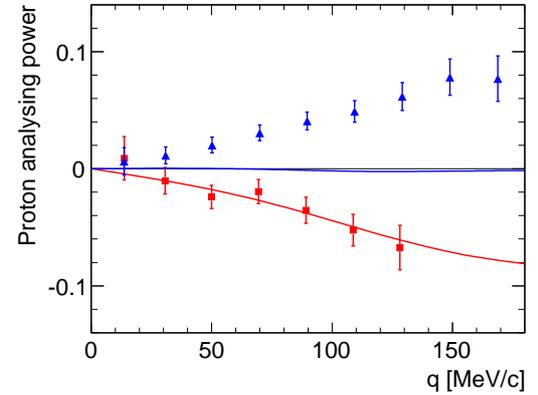}
\caption{\label{fig:AyP} Proton analysing powers $A_{y}^{p}$ for the \dpce\
reaction at $T_d = 1.2$ (red squares) and $2.27$~GeV (blue triangles) for
$E_{pp}<3$~MeV. The error bars include the uncertainties from the target
polarisation. Curves correspond to the theoretical predictions. Note that at
$2.27$~GeV the $A_{y}^{p}$ prediction is very small and hardly visible on
this scale.}
\end{center}
\end{figure}
\begin{figure}[htb]
\begin{center}
\includegraphics[width=0.9\columnwidth]{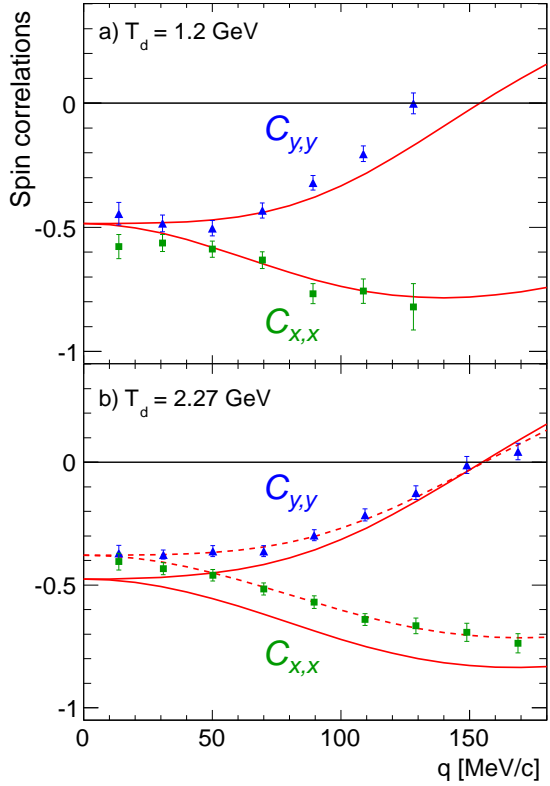}
\caption{\label{fig:Cnn} The spin-correlation coefficients $C_{x,x}$ and
$C_{y,y}$ for the \dpce\ reaction at $T_d = 1.2$ and $2.27$~GeV for
$E_{pp}<3$~MeV. The error bars include the uncertainties from the beam and
target polarisations. The curves are impulse approximation predictions;
dashed curves at $2.27$~GeV correspond to $|\varepsilon(q)|$ being reduced by
$25\%$.}
\end{center}
\end{figure}

In order to extract the spin-correlation parameters from the \dpncep\
reaction, no tensor polarised deuteron beam modes were used (see
Table~\ref{pollist}). In this case the ratio of the polarised $N(q,\phi)$
to unpolarised $N^{0}(q)$ yields has the form~\cite{OHL1972}:
\begin {eqnarray}
\nonumber \frac{N(q,\phi)}{N^{0}(q)} = 1 + Q A_{y}^{p}(q)\cos\phi
+ \fmn{3}{2}P_{z}A_{y}^{d}(q)\cos\phi + \\
+ \fmn{3}{4}P_{z} Q
[(1+\cos2\phi)C_{y,y}(q)+(1-\cos2\phi)C_{x,x}(q)].\phantom{1}
\label{eq:dc_general}
\end {eqnarray}

Although the experiment was designed for the study of spin correlations, by
analysing first the polarised target yields for an unpolarised deuteron beam,
we could obtain the dependence of the target analysing power $A_{y}^{p}$ on
$q$ that is presented in Fig.~\ref{fig:AyP} at 1.2 and 2.27~GeV. This is
predicted very well in impulse approximation at the lower energy but at
2.27~GeV the corresponding prediction can hardly be distinguished from the
$x$-axis on this scale. Equation~(\ref{impulse}) then suggests that there
must be a serious problem also with the SAID determination of the spin-orbit
amplitude $\gamma(q)$ at 1.135~GeV.

In impulse approximation $A_{y}^{d}$ vanishes~\cite{BUG1987},
which is consistent with our measurements at
1.17~GeV~\cite{CHI2009}, and this simplifies the determination
of $C_{x,x}$ and $C_{y,y}$ using data with polarised beam and
polarised target. After binning the normalised counts in
intervals in $q$, the $\cos2\phi$ dependence in
Eq.~(\ref{eq:dc_general}) allowed us to extract the $C_{x,x}$
and $C_{y,y}$ coefficients separately. Note, that the
resolutions in both $E_{pp}$ and $q$ in the cell-target data
are similar to those achieved with the cluster target.

The spin-correlation data for an $E_{pp} < 3$~MeV cut are compared with
theoretical predictions in Fig.~\ref{fig:Cnn}. The good agreement with the
experimental points at $T_d = 1.2$~GeV shows that the two relative phases
between the spin-spin amplitudes are well predicted by the SAID program at
this energy. It can, however, not come as a complete surprise to find that
there are serious discrepancies at $T_d = 2.27$~GeV but, as shown by the
dashed line, these largely disappear if the SAID $\varepsilon(q)$ amplitude
is scaled uniformly by a factor of 0.75, \emph{i.e.}, by the same factor that
brought agreement for the $A_{xx}$ and $A_{yy}$ observables!

\section{Conclusions and Outlook}
\label{sec8}\setcounter{equation}{0} %

We have measured the unpolarised differential cross section and
the Cartesian tensor analysing powers in the \dpcepol\ reaction
for small momentum transfers between the proton and neutron by
using a hydrogen cluster target in combination with a tensor
polarised deuteron beam. The cross section data at 1.2, 1.6,
and 1.8~GeV are very well described in impulse approximation
using the current SAID solution for the $np$ amplitudes. These
amplitudes suffer from much bigger ambiguities at higher
energies and the corresponding cross section prediction is
about 15\% too high compared to our results at 2.27~GeV, though
one must bear in mind the 6\% uncertainty in the overall
normalisation of the data at this energy. The suspicion must
fall on the SAID solution, which predicts an unpolarised $np\to
pn$ cross section that may be up to 10\% too
large~\cite{BIZ1975}, of which the spin-dependent contribution
is also about 10\% too large~\cite{SHA2009}.

The description of $A_{xx}$ and $A_{yy}$ is also very good at the three lower
energies but much poorer at 2.27~GeV. Since one would expect the impulse
approximation to become better as the energy is raised, attention is once
again focussed on the SAID $np$ amplitudes. The strength of
$|\varepsilon(0)|^2$ in $np$ charge exchange relative to
$|\beta(0)|^2=|\delta(0)|^2$ is determined by the spin-transfer parameters
$K_{LL}(0)$ and $K_{NN}(0)$, but there are no measurements of these
quantities in the relevant angular and energy region. This limits severely
the SAID predictive power for the deuteron tensor analysing powers. To fit
our data, we have reduced the SAID prediction for $\varepsilon(q)$ uniformly
by 25\% and this reproduces the results much better. Although this might be
improved further by introducing a $q$-dependence in this factor, the present
data do not justify such a refinement.

By replacing the unpolarised hydrogen cluster-jet target by a polarised
hydrogen gas cell, it was possible to measure the spin-correlation
coefficients $C_{x,x}$ and $C_{y,y}$ in the \dpncep\ reaction with a vector
polarised deuteron beam, but only at 1.2 and 2.27~GeV. The behaviour seen
here is similar to that for the other observables, with a good description
being achieved at 1.2~GeV whereas at 2.27~GeV a reduction of the order of
25\% seems to be required in the $\varepsilon$ input.

As a by-product of the polarised cell experiment, we were also able to
measure the proton analysing power in the reaction. As with the other
observables, impulse approximation reproduces well the small $A_y^p$ signal
at 1.2~GeV but fails completely at 2.27~GeV. This suggests that the SAID
solution for the $\gamma$-amplitude is also unreliable at the higher energy.

In summary, the fact that the impulse approximation with the current SAID
input reproduces well all our data below 1~GeV per nucleon gives us
confidence that the charge-exchange methodology works well. However, the
discrepancies seen at the higher energy can only be resolved by reducing the
strength of the spin-spin amplitudes, especially in the longitudinal
direction, while increasing the spin-orbit contribution. It is therefore
evident that the charge exchange on the deuteron contains valuable
information on the neutron-proton amplitudes. The challenge is to get this
used inside the SAID program.

The experiments reported here were carried out up to the maximum deuteron
energy available at COSY. To go higher in energy at this facility, an
experiment would have to be undertaken in inverse kinematics with a polarised
proton incident on a polarised deuterium gas cell~\cite{GRI2007}, with the
two slow protons being detected in the Silicon Tracking
Telescopes~\cite{SCH2003}. This will allow the studies reported here to be
continued up to 2.9~GeV per nucleon~\cite{KAC2005,CHI2012}.

%
%
\begin{acknowledgement}
We are grateful to the accelerator crew for the reliable operation of COSY
and the deuteron polarimeters. We would like to thank I.~I.~Strakovsky for
many useful discussions and for providing us with values of the current SAID
neutron-proton amplitudes. The work was supported by the COSY FFE programme
and the Shota Rustaveli National Science Foundation Grant 09-1024-4-200.
\end{acknowledgement}

%
%

\end{document}